\documentclass[conference]{IEEEtran}
\usepackage{cite}
\usepackage{amsmath,amssymb,amsfonts}
\usepackage{graphicx}
\usepackage{textcomp}
\usepackage{xcolor}

\begin{document}

\title{Towards Auction-Based Function Placement in Serverless Fog Platforms}

\author{
\IEEEauthorblockN{
David Bermbach\IEEEauthorrefmark{1},
Setareh Maghsudi\IEEEauthorrefmark{2},
Jonathan Hasenburg\IEEEauthorrefmark{1},
Tobias Pfandzelter\IEEEauthorrefmark{1}}
\IEEEauthorblockA{TU Berlin \& Einstein Center Digital Future}
\IEEEauthorblockA{\IEEEauthorrefmark{1}Mobile Cloud Computing Research Group: \texttt{{\{db, jh, tp\}}@mcc.tu-berlin.de}}
\IEEEauthorblockA{\IEEEauthorrefmark{2}Control of Convergent Access Networks Research Group: \texttt{maghsudi@tu-berlin.de}
}
}

\maketitle

\begin{abstract}
The Function-as-a-Service (FaaS) paradigm has a lot of potential as a computing model for fog environments comprising both cloud and edge nodes.
When the request rate exceeds capacity limits at the edge, some functions need to be offloaded from the edge towards the cloud.

In this position paper, we propose an auction-based approach in which application developers bid on resources. This allows fog nodes to make a local decision about which functions to offload while maximizing revenue. For a first evaluation of our approach, we use simulation.
\end{abstract}


\section{Introduction\label{sec:intro}}
Recently, the paradigm of fog computing has received more and more attention. In fog computing, cloud resources are combined with resources at the edge, i.e., near the end user or close to IoT devices, and in some cases also with additional resources in the network between cloud and edge~\cite{paper_bermbach_fog_computing}. While this adds extra complexity for applications, it also comes with three key benefits:
First, leveraging compute resources at or near the edge promises lower response times which is crucial for application domains such as autonomous driving or 5G mobile networks~\cite{paper_bermbach_fog_computing,BonomiFog}.
Second, data can be filtered and pre-processed early on the path from edge to the cloud which reduces the data volume~\cite{paper_pfandzelter_streams_functions}.
Especially in IoT use cases, it is often not feasible to transmit all data to the cloud as the sheer volume of produced data exceeds the bandwidth capabilities of the network~\cite{paper_zhang_cloud_is_not_enough_GDP}. Third, keeping parts of applications and data at the edge can help to improve privacy, e.g., by avoiding centralized data lakes in the cloud~\cite{paper_pallas_fog4privacy,paper_grambow_fog_video_surveillance,paper_bermbach_fog_computing}. Overall, fog computing, thus, combines the benefits of both cloud and edge computing.

While there are many open research questions in fog computing, a key question has not been answered yet: Which compute paradigms will future fog applications follow? In previous work~\cite{paper_bermbach_fog_computing}, we argued that a serverless approach -- under which we understand Function-as-a-Service (FaaS) -- will be a good fit for the edge. Our reasoning for this is that resources at the edge are often very limited so that provisioning them in small function slices is much more efficient than provisioning them as virtual machines or long-running containers. Additionally, the idea of having strictly stateless functions separated from data management~\cite{paper_baldini_openwhisk,paper_hellerstein_serverless} is also a key requirement for easily moving parts of applications between edge and cloud.

Now, assuming a fully serverless world in which application components run as functions on FaaS platforms in the cloud, at the edge, and possibly in medium-sized data centers in the network in between, the question of how to distribute fog application components can be reduced to the issue of function placement across multiple geo-distributed sites. In this position paper, we propose an approach for this kind of function placement. Hence, we make the following contributions:
\begin{enumerate}
	\item We describe an approach that uses a decentralized auction scheme to control function placement (section~\ref{sec:approach}).
	\item We present a simulation tool and the underlying system model that allows to study the effects of different parameters on such an auction-based placement approach (section~\ref{sec:simulation}).
	\item We discuss a number of insights gained from simulation runs (section~\ref{sec:eval}).
	\item We identify and discuss future research directions (section~\ref{sec:discussion}).
\end{enumerate}

\section{Background\label{sec:background}}
In this section, we will briefly introduce and describe our understanding of FaaS, fog computing, and auctions. The remainder of this paper will be based on this terminology.

\subsection{Function-as-a-Service (FaaS)} \label{subsec:faas}
For our purposes, FaaS is a simple (cloud) service model in which application developers deploy the code of a single function on an execution platform. The platform directs incoming requests to that function, auto-scales the function ``executor'' (often Docker containers) as needed, and bills the developers based on actual usage. Key characteristics are stateless functions leveraging external services, particularly storage, with limited execution durations. This corresponds to the service models of, e.g., AWS Lambda\footnote{aws.amazon.com/lambda} or OpenWhisk~\cite{paper_baldini_openwhisk}.

\subsection{Fog Computing}
There are multiple competing definitions for the term ``fog computing''. Sometimes, it is used as a synonym for edge computing, sometimes it is used to refer to compute resources between edge and cloud, and sometimes it is used to refer to everything beyond the cloud (possibly including it) down to (embedded) device level.
For our purposes, we will follow the definition of~\cite{paper_bermbach_fog_computing}, i.e., fog computing comprises resources at the edge, in the cloud, and all resources in the network between edge and cloud. To the latter, we refer to as ``intermediary nodes'' or short ``intermediary''. In a mobile network scenario, ``the edge'' would be compute resources that are collocated with the cellular towers, ``the cloud'' would be the offerings of providers such as AWS\footnote{aws.amazon.com}, and ``intermediaries'' would be small- to medium-sized data centers within the network of the Internet provider.

\begin{figure}[t]
    \centering
    \includegraphics[width=\columnwidth]{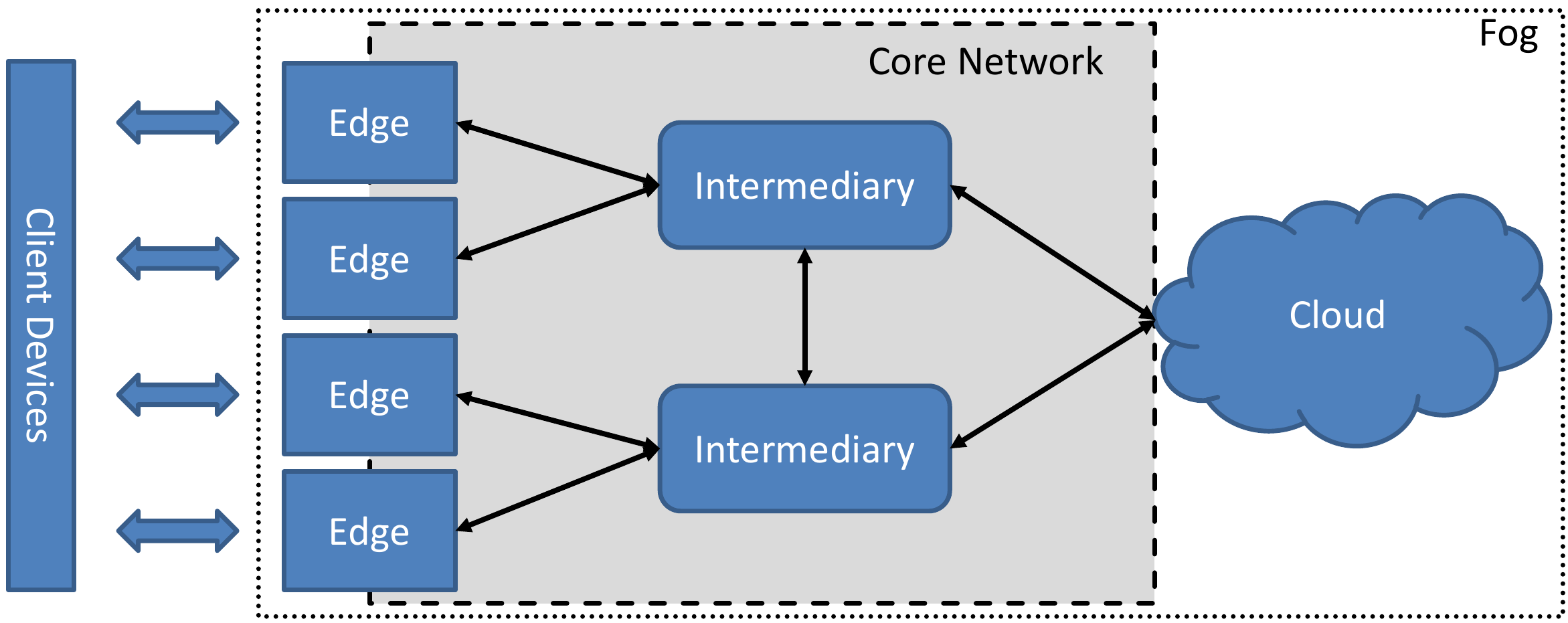}
    \caption{Overview of Cloud, Edge, and Fog Computing (adapted from~\cite{paper_bermbach_fog_computing})}
    \label{fig:fog}
\end{figure}

\subsection{Fog-Based FaaS Platforms}
Beyond commercial FaaS platforms provided as cloud service, e.g., AWS Lambda or IBM Cloud Functions\footnote{cloud.ibm.com/functions}, systems such as OpenWhisk~\cite{paper_baldini_openwhisk} (the system behind IBM Cloud Functions) or OpenLambda~\cite{paper_hendrickson_openlambda} have been made available as open source. For a fog deployment, such systems are a good fit for cloud nodes but also for larger intermediary nodes. Moreover, there is LeanOpenWhisk\footnote{github.com/kpavel/incubator-openwhisk/tree/lean} which is a good fit for intermediary or larger edge nodes. As we have seen in experiments, LeanOpenWhisk is not yet ``lean'' enough for smaller edge nodes such as a single machine or even a Raspberry Pi; tinyFaaS~\cite{paper_pfandzelter_tinyfaas} can be an alternative for smaller edge nodes. Overall, this shows that the necessary systems for deployment in the fog already exist as open source (or will be there shortly).

\subsection{Auctions\label{sec:auctions}}
Auction is a way to sell a commodity at some price that is determined through the competition among potential buyers (bidders). 
%
%
The seller (auctioneer) sets the auction's rule, which can be roughly written as the pair $(R,P)$, with
$R$ being the allocation rule that determines the winner among the bidders, and
$P$ being the payment rule that describes the price every participant has to pay to the seller.

While competing, the bidders have to follow the auction's rules set by the seller (auctioneer). 
There is a large body of auction rules. This includes, but is not limited to
\begin{itemize}
\item \textbf{First/Second-price Sealed-bid Auction:} The object is awarded to the highest bidder, at the price of its bid (first-price) or the bid of the second-highest bidder (second-price), which ensures truthful bids.
\item \textbf{English Auction:} The price is raised continuously by the bidders. The winner is the last bidder to remain and pays an amount equal to the price at which all other bidders have dropped out of the auction.
\item \textbf{Dutch Auction:} An initial price is set, and it is lowered continuously by the auctioneer. The winner is the first bidder to agree to pay any price.
\item \textbf{All-Pay Auction:} All bidders submit bids, and the high bidder wins, but every bidder has to pay its bid.
\item \textbf{Penny Auction:} The price is raised continuously by the auctioneer, and the winner pays the current bid. But every bidder pays for all bids along the way, i.e., bidding costs something.
\end{itemize}
In designing and evaluating the auction rule, the desiderata are vast. One of the most important characteristics is truthfulness or incentive-compatibility, which means that for each bidder the best strategy is to bid its value regardless of the strategies of other bidders. Other factors include revenue, participation, or fairness.

\section{Auction-Based Function Placement\label{sec:approach}}
%
%
While a number of FaaS systems is already available, as discussed in section~\ref{subsec:faas}, it is still not clear how to connect different FaaS deployments.
Ideally, we would want edge, intermediary, and cloud deployments to coordinate function placement among each other.
For instance, a request arriving in the cloud should probably still be processed in the cloud. A request arriving at the edge, however, should in most cases (exceptions include functions that require data located elsewhere~\cite{paper_hellerstein_serverless}) be executed right at the edge. Only when the load on the edge node exceeds the available capacity should the request be delegated to the intermediary (and likewise from the intermediary to the cloud). This concept, similar to cloud bursting, is based on the intuition of the cloud providing (the illusion of) infinite resources~\cite{nist_definition_cloud_computing}.

Based on this, we can conclude that function placement is straightforward when nodes have spare capacity (see also~\cite{paper_pfandzelter_streams_functions} for a more general discussion) but becomes challenging when, for instance, an edge node is overloaded. In such a situation, the question is which request should be delegated to the next node on the path to the cloud as that request will incur extra latency. For this decision, we can imagine a number of criteria, e.g., to prefer short-running functions over long-running ones (and vice versa), to consider bandwidth impacts, to give some clients preference over others, or to prefer latency-critical functions over less critical ones.

In contrast to these, we propose to use an auction-based approach which has been proven in multiple domains (e.g.,~\cite{Huang08:ABR,Gao11:MAP,Zhao14:HCT,Yang16:IMC,Kantarci14:TSP,Zhang14:DRP,Chandrashekar07:ABM,Jayaweera11:ACC}) to lead to an efficient resource allocation: When application developers deploy their function to an integrated fog FaaS platform, they also attach two bids to the executable. The first bid is the price that the respective developer is willing to pay for a node to store the executable (in \$/s), the second bid is for the execution of the function (in \$/execution). In practice, both bids are likely to be vectors so that developers could indicate their willingness to pay more on edge nodes than in the cloud or even on a specific edge node. We explicitly distinguish bids for storage of the executable and the execution as edge nodes might encounter storage limits independent from processing limits. Both auctions closely resemble a sequence of first-price sealed-bid auctions discussed in section~\ref{sec:auctions}. Please note two things: (i) In practice, a second-price auction would be preferable.
For ease of explanation, however, we use the first-price auction in the following as it makes no difference in our examples. (ii) As there is likely to be a minimum price equivalent to today's cloud prices, we consider both bids a surcharge on top of the regular cloud price. For ease of explanation, we again ignore this base price in the following.

\textbf{Storage Bids:}
The nodes in our approach -- edge, intermediaries, and cloud -- analyze these bids and can make a local decision, i.e., act as auctioneers, which is important for overall scalability. We assume that cloud nodes will accept all bids that exceed some minimum price (see the discussion of base prices above). All other nodes will check whether they can store the executable. When there is enough remaining capacity, nodes simply store the executable together with the attached bids and start charging the application based on the storage bid. When there is not enough disk space left, nodes decide whether they want to reject the bid or remove another already stored executable. For this, nodes try to maximize their earnings. A very simple strategy for this, comparable to standard bin packing, is to order all executables by their storage bid (either the absolute bid or the bid divided by the size of the executable) and keep removing executables until the new one fits in. Of course, arbitrary complex strategies can be used and could even consider the processing bid and the expected number of executions. In the end, this leads to a situation where some or all nodes will store the executable along with its bids.

\textbf{Processing Bids:}
When a request arrives, nodes have to decide whether they \emph{want} to process said request. A node \emph{can} process a request when it stores the corresponding executable and when it has sufficient processing capacity to execute the function. We can imagine arbitrarily complex schemes for making the decision on whether to execute the request or not. For instance, nodes might try to predict future requests (and decide to wait for a more lucrative one) or they might try to queue a request shortly -- all with the goal of maximizing earnings. In fact, we believe that this opens interesting opportunities for future work.

For the explanations in this paper, we will assume that a node will decide to process a request as soon as it is capable of doing so. This means that upon receipt of a set of incoming requests, a node will follow these four steps:
\begin{enumerate}
	\item Reject all requests for which the executable is not stored locally.
	\item Sort remaining requests by their bid (highest first) into a request list.
	\item While capacity is left, schedule requests from request list.
	\item Reject all requests that exceed the capacity.
\end{enumerate}

Rejected requests are then pushed to the next node on the path to the cloud. As the cloud stores by definition all executables and has for all practical purposes unlimited capacity, all requests will at least be served in the cloud.

Overall, this approach solves the question of function allocation, all functions are executed as close as possible to the edge; under high load, it will also have the effect that prices increase towards the edge. Of course, it is possible to identify scenarios where this simple allocation scheme leads to the opposite effect. For instance, when long-running requests with low bids are directly followed by requests with high bids, the overall result may well be that all high-bid requests are served in the cloud. This, however, is a corner case scenario and we leave the identification of both optimal bidding strategies as well as optimal node decision rules for future work.

\section{Model and Simulation Approach\label{sec:simulation}}
To evaluate our approach, we have implemented a simulation tool in Kotlin which is available on GitHub\footnote{https://github.com/dbermbach/faas4fogsim}. In the tool, we distinguish three types of nodes: edge, intermediary, and cloud. All three node types have a distinct storage and processing capacity, we assume that all requests get the same compute power but may take arbitrarily long, i.e., the processing capacity is specified as the number of requests which can be handled in parallel. All nodes treat bids in the way described in section~\ref{sec:approach}. The only difference is that a simulation run has a fixed duration in time units: if a non-cloud node cannot complete a request before the simulation ends, it will also push it towards the cloud. For future research on different auction strategies, it is only necessary to change the methods \texttt{offerExecutable} and \texttt{offerRequests} in ComputeNode.kt which implement the respective allocation rules (see section~\ref{sec:background}).

Beyond the parameters already mentioned, users can specify the node setup (number and type of nodes and their interconnection), average latency for request execution as well as latency between edge and intermediary or intermediary and cloud, average storage and processing bids, number and average size of executables, and the number of requests that should arrive at each edge node. While the simulation tool can handle more diverse configurations in its calculations, we recommend to stick with a few standard settings so that users only need to modify a configuration object (\texttt{Config} in Simulator.kt). In these settings, we make the assumption that all requests will arrive at the edge and that there are at least one edge, intermediary, and cloud node each. Our implementation also assumes that bids are identical across all node types.

To support reproducibility of simulation runs, the tool explicitly sets the random seed; parameters specified as average \textit{X} follow a uniform distribution over a specified range.
\begin{figure}[t]
    \centering
    \includegraphics[width=0.9\columnwidth]{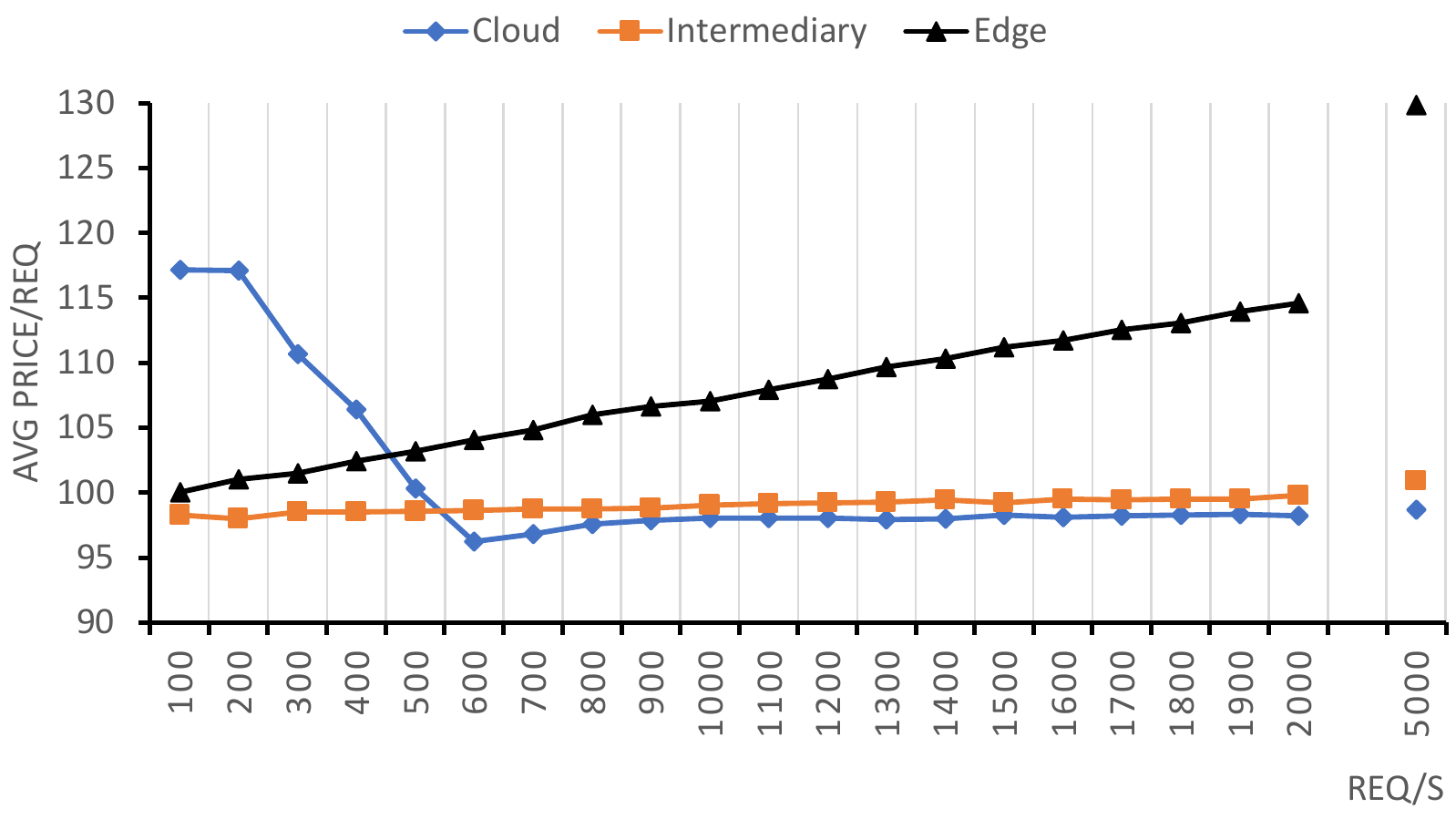}
    \caption{Experiment 1: Average Price per Function Execution across Edge, Intermediary, and Cloud}
    \label{fig:exp1-earnings}
\end{figure}

\begin{figure}[t]
    \centering
    \includegraphics[width=0.93\columnwidth]{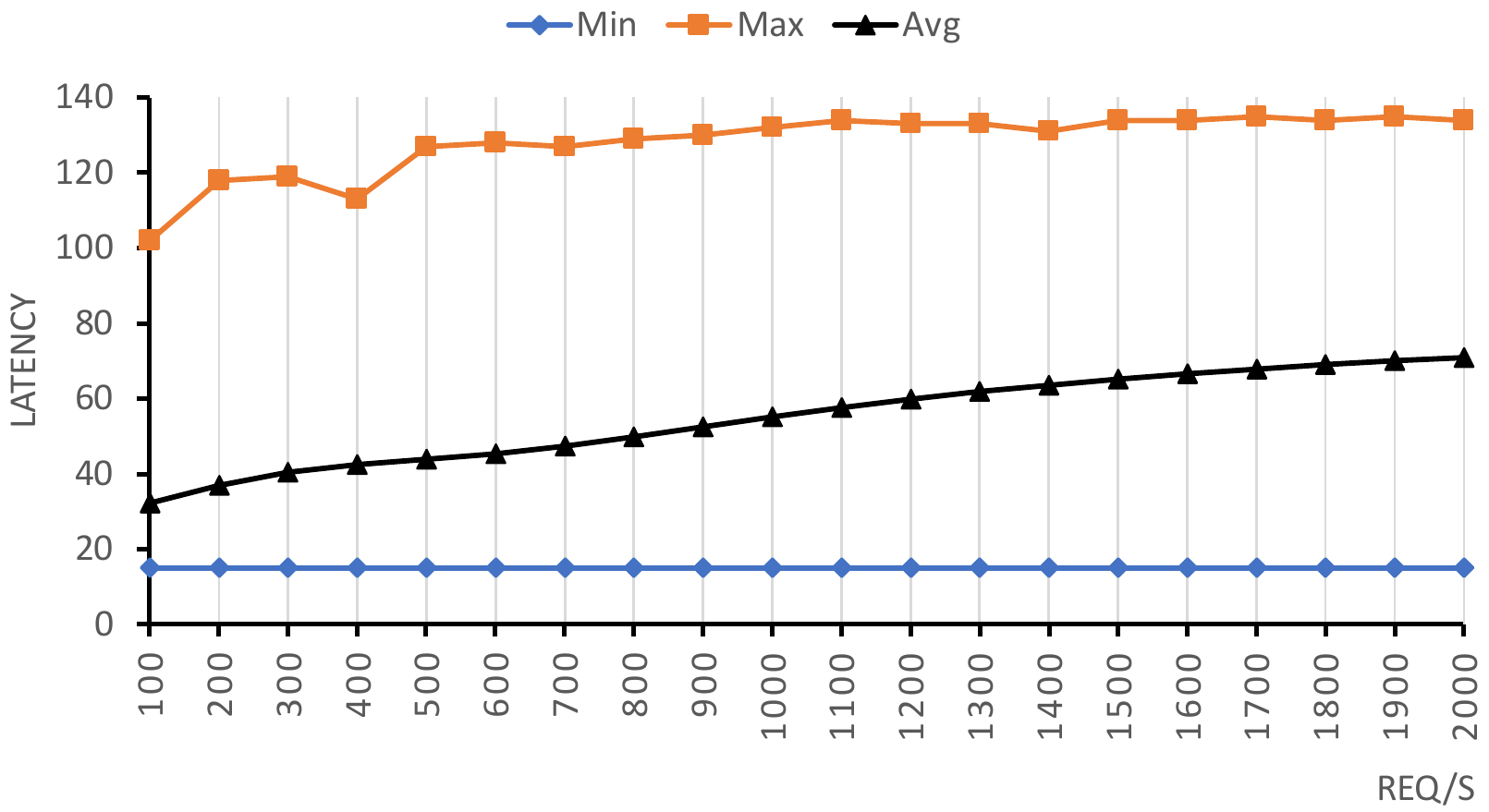}
    \caption{Experiment 1: Overall Request Latency under Varying Load}
    \label{fig:exp1-latency}
\end{figure}

\section{Insights from Simulation\label{sec:eval}}

In this section, we describe insights we gained from simulation. Namely, we ran two different simulation experiments: The first analyzed the effect of processing prices on function placement and latency depending on the request load. The second studied the effect of storage prices on function placement and latency depending on the demand for storage.

\subsection{Configuration}
For both simulation experiments, we had one edge, intermediary, and cloud node and simulated a period of two minutes; we also set the average processing latency as 30ms, the average edge to intermediary latency as 20ms, and the average intermediary to cloud latency as 40ms.

In the first experiment, we chose storage parameters in a way that they would not influence the experiment at all and only limited the processing capabilities on the edge and the intermediary. In our parameter settings, anything up to 166req/s could on average be handled at the edge and anything up to an average of 832req/s at edge and intermediary. Anything beyond this required the cloud to handle the load. Each request had a random processing bid from the interval [50;150] (uniformly distributed). We repeated this experiment with varying request load: from 100 to 2000req/s in steps of 100. Our experiment, thus, simulated 12,000 to 240,000 requests. We also ran a single simulation round for 5000req/s to verify trends.

In the second experiment, we chose processing parameters that should not affect the simulation result. Edge nodes had a storage capacity of 100, intermediaries of 500, and the cloud was unlimited; the average size of an executable was 10 and came with random storage bids from the interval [50;150] (uniformly distributed). We repeated this experiment with varying numbers of executables starting with 5 (which on average should still all fit in the edge node), over 50 (on average edge nodes store 20\% of all executables and intermediaries still store all of them), up to 100 in steps of 5.
For all other parameters, we chose sensible settings that were identical in both experiments; the precise settings can be found in the GitHub repository of our tool.

\subsection{Experiment 1: Effect of Processing Prices}

As can be seen in figures~\ref{fig:exp1-earnings} and~\ref{fig:exp1-latency}, the simulated system is lightly loaded up to about 500req/s. Here, the random distribution of request generation plays a much bigger role than the auction mechanism. For the earnings, we can see in figure~\ref{fig:exp1-earnings} that the average price for executing a function at the edge gradually increases. For the intermediary it also increases continuously (from 98.3 at 100req/s to 99.8 at 2000 req/s) but at a much smaller scale since the edge is not yet overloaded enough to delegate requests with higher bids. The price range for bids is capped at 150 so that the edge is already fairly close to this maximum at 5000req/s.
For latency, we can see that it linearly increases with the load as more and more requests (the lower paying ones) are no longer served on the edge but rather delegated to intermediary and cloud.

Overall, this experiment shows that even the simplistic auction scheme that we chose is indeed an efficient mechanism to handle function allocation across fog nodes with the goal of determining the placement based on payment preferences of application developers.

\subsection{Experiment 2: Effect of Storage Prices}

\begin{figure}[t]
    \centering
    \includegraphics[width=0.9\columnwidth]{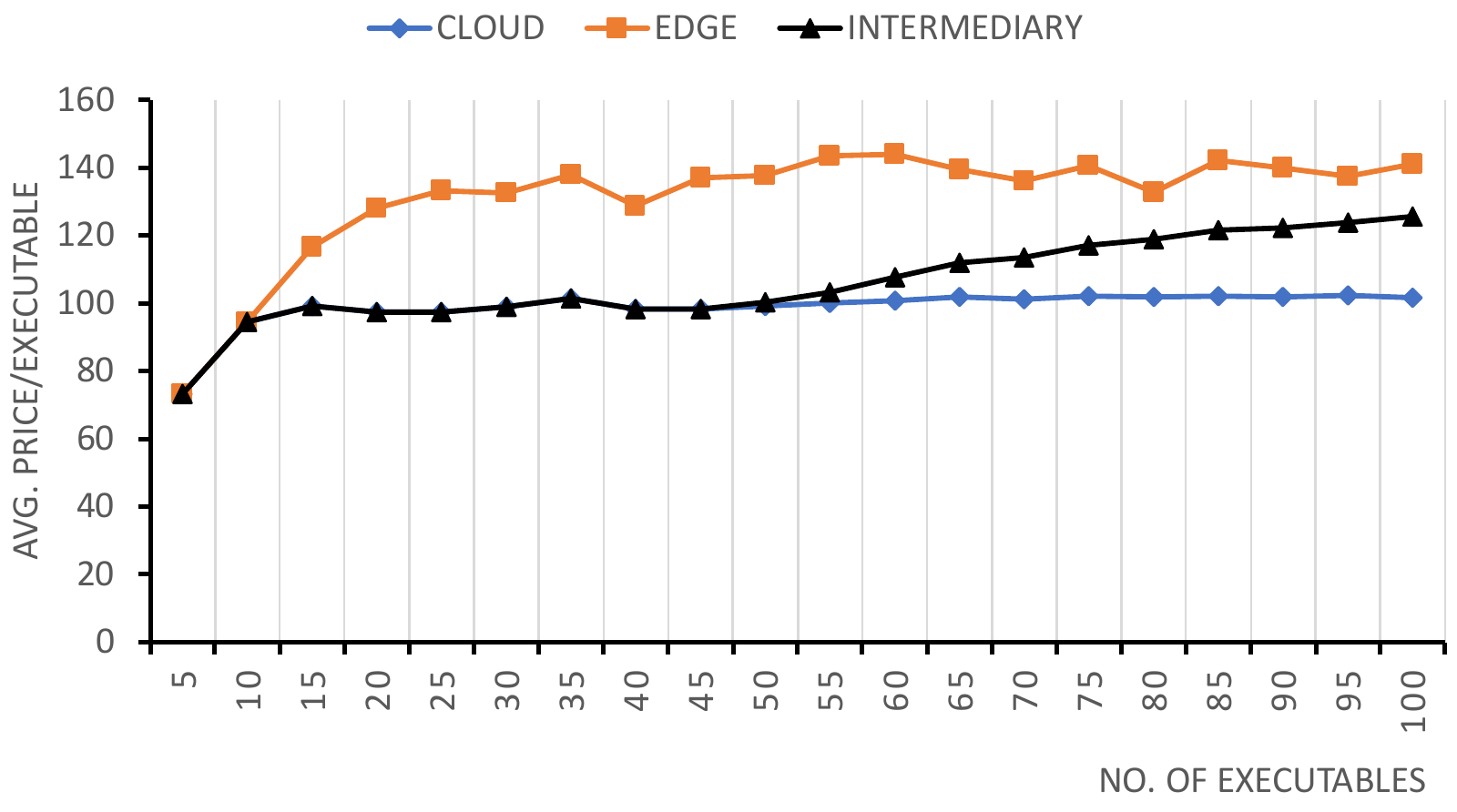}
    \caption{Experiment 2: Average Price per Stored Function across Edge, Intermediary, and Cloud}
    \label{fig:exp2-earnings}
\end{figure}

\begin{figure}[t]
    \centering
    \includegraphics[width=0.9\columnwidth]{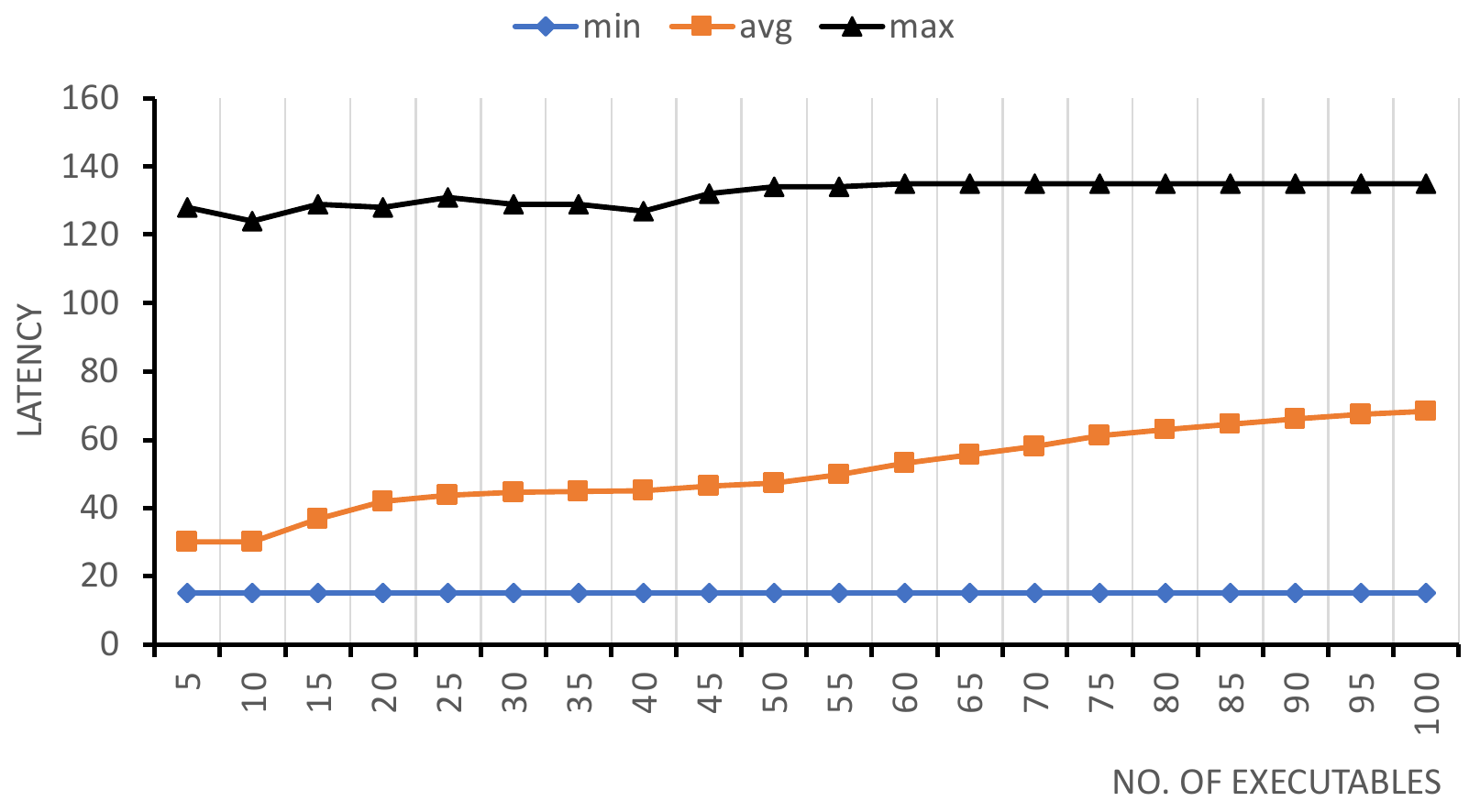}
    \caption{Experiment 2: Overall Request Latency at Varying Availability of Function Executables}
    \label{fig:exp2-latency}
\end{figure}

As can be seen in figure~\ref{fig:exp2-earnings}, which shows the average storage price of all executables stored on a node, the edge does not have any choice of executables when only five are offered. Likewise, the intermediary only starts to evict executables once it exceeds its capacity of about 50. Afterwards, both curves gradually increase towards the upper limit of 150 which, however, is not reached in this experiment as the randomization means that only very few executables will be offered at a storage bid of 150. In addition, the average storage price in the cloud is about 100 which is as expected as the cloud accepts all executables in our simulation setting.

Figure~\ref{fig:exp2-latency}, in contrast, shows how this leads to increasing latency across requests: With every additional executable, the share of executables available at the edge (or on the intermediary) decreases so that requests are increasingly delegated towards the cloud since the function can, without an executable, not be run.

Overall, this experiment shows that our auction-based approach is an efficient mechanism to determine distribution of binaries across a fog network when there is a lack of storage capacity.

\section{Discussion and Future Work\label{sec:discussion}}
With our simulation model, we showed that it is indeed feasible to address function placement for fog-based FaaS platforms based on distributed auctions. In our approach, however, the bids and the auction rules follow a rather simplistic approach. We believe that this opens up interesting opportunities for future research:

First, application developers could specify bids in a more intelligent way, considering the location of edge nodes and their clients as well as availability of resources, i.e., they should provide high bids for scarce resources that are urgently needed while offering low bids, e.g., for processing and storage in the cloud. This also opens up a few game-theoretic questions about combinations of storage and processing bids, e.g., to use a high storage bid with a bloated executable to get a single tenant edge node combined with a very low processing bid to reduce cost.

Second, our current nodes are rather simplistic in that they do not look into the future and do not couple storage and processing bid -- they simply accept all executables that fit in and serve whatever comes in ordered by the bids. While this is in line with the chosen auction model, much higher revenue could be achieved based on more intelligent decisions. For instance, just considering the processing bid as well as past invocation numbers coupled with the storage bid would result in more profitable situations. As another example, it may make sense to delegate a processing request even though the node is not saturated when a periodic request with a higher processing cost is expected to arrive shortly.

Combining both aspects, significant efficiency improvements are possible when application developers try to minimize cost while nodes try to maximize earnings. Beyond this, there are of course a number of technical challenges in implementing such an integrated platform, e.g., for dealing with failures. We believe that both the game-theoretic challenges as well as the systems challenges are worthy of further research attention. Finally, there is also the aspect of co-managing data placement~\cite{paper_hasenburg_towards_fbase,techreport_hasenburg_2019,paper_zhang_cloud_is_not_enough_GDP,paper_confais_ipfs4fog} and function placement~\cite{paper_pfandzelter_streams_functions,hasenburg_fogexplorer_2018,paper_baresi_serverless4fog
} which is a promising avenue to pursue~\cite{paper_hellerstein_serverless,sreekanti2020cloudburst}.

Beyond this, our approach obviously takes a rather particular approach as application developers will get no guarantees at all where their function will be executed -- the placement of functions only depends on the amount of the bids at a specific point in time. There is, hence, no way to guarantee bounds on response time and other quality dimensions, i.e., the infrastructure provider does not provide any guarantees beyond \emph{eventually} executing every function \emph{somewhere}. In a way, this is comparable to airline seat assignment schemes in which economy class passengers can bid for business class upgrades. While this is undesirable from an application developer perspective, our proposed approach has the benefit that it requires no centralized coordination at all, i.e., it scales well, and that it is likely to maximize earnings for the infrastructure provider. Also, the lack of guarantees is perfectly inline with the state-of-practice in current cloud services which provide minimal to no guarantees at all as part of their SLAs~\cite{bradshaw2011contracts}. Nevertheless, if our proposed approach were used in practice, we would expect infrastructure providers to let customers reserve subsets of their resources at a fixed price -- comparable to 5G slicing~\cite{foukas2017network} -- hence using the proposed auctioning scheme only for a subset of their resources. This would lead to the interesting situation that application developers that actually need quality guarantees can pay for it (comparable to booking a business class seat) while others can still bid on the remaining resources (comparable to bidding on a business class upgrade).

\section{Related Work\label{sec:relwork}}
While auctions have been used in multiple domains, e.g.,~\cite{Huang08:ABR,Gao11:MAP,Zhao14:HCT,Yang16:IMC,Kantarci14:TSP,Zhang14:DRP,Chandrashekar07:ABM,Jayaweera11:ACC}, we are not aware of any directly related work in this regard. Other approaches, e.g.,~\cite{paper_nastic_serverless_at_edge}, argue for function placement in the fog based on global optimization; later work by Rausch et al.~\cite{paper_rausch_towards_serverless4edge} show through experiments that centralized schedulers such as the Kubernetes scheduler cannot handle the load necessary for function placement in a fog-based FaaS. According to the authors, there is currently no solution which is able to handle such scheduling across edge nodes and clouds.

Other approaches from mobile edge computing, e.g.~\cite{paper_machen_live_migration_mec,paper_wang_mec_service_migration} propose to use long-running services that are live migrated upon user movement instead of provisioning small functions when needed. While this is certainly an alternative, we believe that it is not a particularly efficient use of scarce edge resources as long-running services are likely to be idle most of the time while still blocking resources.

Baresi and Medonca~\cite{paper_baresi_serverless4fog} propose independent FaaS platform instances distributed over the fog which offload functions to other instances when needed. They propose to place functions based on current load and request requirements. Unfortunately, the authors are not particularly clear about how and where their system makes these function allocation decisions: Apparently, requests are first routed to a regional load balancer (comparable to our intermediary nodes) which then distributes requests. It remains unclear whether there is more than one regional load balancer and how the distribution decision is actually made. In either case, sending a request from the edge to an intermediary first before processing it at the edge adds unnecessary latency.

Commercial solutions such as AWS Greengrass\footnote{aws.amazon.com/greengrass} provide capabilities to execute functions at the edge and link them to the cloud. In contrast to our proposed approach, functions at the edge and in the cloud need to be addressed in different ways so that the choice of execution location is left to the application developer.

Finally, there are a number of alternative strategies for function placement as, e.g., described in a survey on offloading techniques by Aazam et al.~\cite{paper_aazam_offloading_survey}, in which quality of service constraints are considered for identifying an ``optimal'' placement~\cite{basic2019fuzzy,de2019multi}.

\section{Conclusion\label{sec:concl}}
Serverless FaaS is a promising paradigm for fog environments.
In practice, fog-based FaaS platforms have to schedule the execution of functions across multiple geo-distributed sites, especially when edge nodes are overloaded.
Existing approaches mostly argue for centralized placement decisions which, however, do not scale~\cite{paper_rausch_towards_serverless4edge}.

In this position paper, we proposed to follow an auction-based scheme in which application developers specify bids for storing executables and executing functions across the fog.
This way, overloaded fog nodes can make local decisions about function execution so that function placement is no longer a scalability bottleneck.
Furthermore, we showed through simulation that our approach asserts that all requests are served while maximizing revenue for overloaded nodes. We argue for further research on intelligent agents and systems in this area.

%
\bibliographystyle{IEEEtran}
\bibliography{cites}

\end{document}